# Detection of Ocean Glint and Ozone Absorption Using LCROSS Earth Observations

Tyler D. Robinson[1,6,7], Kimberly Ennico[2], Victoria S. Meadows[3,6,7],
William Sparks[4,6,7], D. Ben J. Bussey[5,7], Edward W. Schwieterman[3,6,7],
and Jonathan Breiner[3,6]

[1] NASA Ames Research Center, MS 245-3, Moffett Field, CA 94035, USA; tyler.d.robinson@nasa.gov
[2] NASA Ames Research Center, MS 245-6, Moffett Field, CA 94035, USA
[3] Astronomy Department, University of Washington, Seattle, WA 98195, USA
[4] Space Telescope Science Institute, Baltimore, MD 21218, USA
[5] Johns Hopkins University Applied Physics Laboratory, Laurel, MD 20723, USA
[6] NASA Astrobiology Institute's Virtual Planetary Laboratory
[7] NASA Lunar Science Institute

ABSTRACT

The Lunar CRater Observation and Sensing Satellite (*LCROSS*) observed the distant Earth on three occasions in 2009. These data span a range of phase angles, including a rare crescent phase view. For each epoch, the satellite acquired near-infrared and mid-infrared full-disk images, and partial-disk spectra at 0.26–0.65 μm ($\lambda/\Delta\lambda$ ~500) and 1.17–2.48 μm ($\lambda/\Delta\lambda$ ~50). Spectra show strong absorption features due to water vapor and ozone, which is a biosignature gas. We perform a significant recalibration of the UV-visible spectra and provide the first comparison of high-resolution visible Earth spectra to the NASA Astrobiology Institute's Virtual Planetary Laboratory three-dimensional spectral Earth model. We find good agreement with the observations, reproducing the absolute brightness and dynamic range at all wavelengths for all observation epochs, thus validating the model to within the ~10% data calibration uncertainty. Data-model comparisons reveal a strong ocean glint signature in the crescent phase dataset, which is well matched by our model predictions throughout the observed wavelength range. This provides the first observational test of a technique that could be used to determine exoplanet habitability from disk-integrated observations at visible and near-infrared wavelengths, where the glint signal is strongest. We examine the detection of the ozone 255 nm Hartley and 400–700 nm Chappuis bands. While the Hartley band is the strongest ozone feature in Earth's spectrum, false positives for its detection could exist. Finally, we discuss the implications of these findings for future exoplanet characterization missions.

# 1    INTRODUCTION

Earth is our only example of a world capable of maintaining liquid water on its surface, and it will always be the best-studied example of a habitable planet. Thus, when attempting to understand techniques for detecting and characterizing habitable exoplanets, we use Earth as our archetype. Due to the success of NASA's *Kepler* mission (Borucki et al. 2003; Basri et al. 2005), we are discovering increasingly smaller worlds (Borucki et al. 2010; Borucki et al. 2011a; Borucki et al. 2011b; Batalha et al. 2013), and it is becoming clear that potentially habitable worlds may be common (Traub 2012; Dressing & Charbonneau 2013; Kopparapu 2013).

It is a high priority goal in exoplanet research to obtain and interpret the spectrum of a potentially habitable planet, to characterize that world for habitability, and to search for signs of life on its surface or in its atmosphere (Des Marais et al. 2002; Seager et al. 2005; Meadows 2006). To support this goal, several studies to interpret spectroscopic and photometric observations of the distant Earth have been undertaken. Sagan et al. (1993) used observations of Earth by the *Galileo* spacecraft to characterize Earth's surface and atmosphere, and made several arguments for the evidence of life on the planet. However, that study relied on spatially resolved imaging and spectroscopy, which will not be achievable for exoplanets in the near future, thus complicating interpretation. More recently, observations of the distant Earth from NASA's *EPOXI* mission (Livengood et al. 2011) have been used to characterize and map the distribution of different surfaces using rotational variation from spatially unresolved photometric observations (Cowan et al. 2009; Cowan et al. 2011; Fujii et al. 2011; Cowan & Strait 2013). Additionally, variability in absorption features of $H_2O$ and $O_2$ in near-infrared (NIR) spectra from the *EPOXI* mission were used to argue that these species have different vertical and horizontal distributions in Earth's atmosphere (Fujii et al. 2013).

Studies aimed at interpreting observations of the unresolved Earth (the "Pale Blue Dot") are hindered, in part, by a paucity of data suitable for such work. While an armada of satellites continuously monitors our planet, their instruments typically do not obtain a simultaneous view of the entirety of Earth's disk. Thus, these data are not appropriate for work aimed at studying Earth as an exoplanet, although it is possible to weave together these observations into a disk-integrated product (Hearty et al. 2009; Gómez-Leal et al. 2012). Ground based observations of Earth's reflected light from the lunar night side (Danjon 1928), or Earthshine, have proven useful for studying Earth's albedo (Goode et al. 2001; Pallé et al. 2003; Qiu et al. 2003) and for identifying certain features in Earth's spectrum (Arnold et al. 2002; Turnbull et al. 2006; Sterzik et al. 2012), but are seldom calibrated on an absolute scale and are limited to wavelengths where Earth's atmosphere is not opaque. Spacecraft observations of the distant Earth (Sagan et al. 1993; Christensen & Pearl 1997; Livengood et al. 2011) are most analogous to what future exoplanet characterization missions may observe, but are very rare.

Given the difficulties associated with obtaining data suitable for studying Earth in the context of an exoplanet, models are instead used to simulate photometric and spectroscopic observations of the Pale Blue Dot. There now exists a hierarchy of one-dimensional and three-dimensional spectral models, spanning a range of approaches to simulating the spectrum of an unresolved Earth. One-dimensional models typically simulate disk-integrated spectra of Earth by treating the atmosphere as a collection of plane-parallel slabs (Woolf et al. 2002; Kaltenegger et al. 2007; Stam 2008; Rugheimer et al. 2012), and must employ certain assumptions or simplifications to reduce the full three-dimensional complexity of Earth into a single dimension. To accomplish this, it is common to use either the average Earth atmospheric composition, temperature, and cloud and surface coverage, or to implement a weighting scheme for these parameters based on the particular viewing geometry of the model (Montañés-Rodríguez et al. 2006).

Three-dimensional spectral models consider longitudinal, latitudinal, and, sometimes, vertical variations in Earth's atmospheric and surface state. The relatively high computational cost associated with treating such variations has incentivized the development of a range of complexities in three-dimensional spectral Earth models. Some models simulate only the reflection of sunlight from clouds and Earth's surface, and ignore atmospheric absorption and emission. These three-dimensional reflectance models are most appropriate at visible wavelengths, and have been used to study rotational variability in Earth's reflected light spectrum (Ford et al. 2001), and the ability of ocean glint, or the mirror-like reflection of sunlight off Earth's oceans, to increase our planet's reflectivity at crescent phases (Williams & Gaidos 2008). Some three-dimensional reflectance models (Pallé et al. 2003; Oakley & Cash 2009; Fujii et al. 2010; Sanromá & Pallé 2011) specify the angular-dependent reflectivity of clouds and surfaces using so-called scene models (e.g., Manalo-Smith et al. 1998), which have been designed to match, or are derived from, satellite observations.

A final category of three-dimensional spectral models uses a numerical radiative transfer model to compute the wavelength-dependent brightness of small patches of Earth, and then sums these together to provide a disk-integrated product (Tinetti et al. 2006; Robinson et al. 2010; Fujii et al. 2011; Robinson et al. 2011; Sanromá et al. 2013). Using this technique allows these models to simulate Earth's appearance across a wide range of wavelengths, spectral resolutions, and viewing geometries. In particular, such models offer superior predictions of Earth's disk-integrated brightness at wavelengths where atmospheric gases absorb. Additionally, these higher complexity models are more accurate in their treatment of Earth's phase-dependent brightness, which is dominated by Rayleigh and cloud scattering, and by specular reflectance from the ocean surface (i.e., glint).

To be a trustworthy stand-in for real data, it is important that all spectral Earth models be rigorously validated against observations. This validation is critical for preventing biases and incorrect predictions when attempting to understand characterization techniques for Earth in the context of an exoplanet. Some of the models discussed above have been compared to calibrated snapshot observations of Earth (Tinetti et al. 2006; Rugheimer et al. 2012), which can still allow errors and biases to persist. For example, while the cloud treatment in a spectral model can be tuned to reproduce a snapshot observation (Tinetti et al. 2006), these tuning parameters may not be valid at other phases or times, thus leading to inaccurate brightness predictions (see, e.g., discussion in Robinson et al. 2011). Only a very limited number of spectral Earth models have been validated against calibrated, temporally resolved photometric (Fujii et al. 2011) and spectroscopic (Robinson et al. 2011) observations of Earth.

On three occasions in 2009, NASA's Lunar CRater Observation and Sensing Satellite (*LCROSS*) (Colaprete et al. 2012; Ennico et al. 2012) acquired high spectral resolution ultraviolet (UV) and visible observations, as well as low spectral resolution NIR observations and mid-infrared (MIR) images of Earth. Most significantly, *LCROSS* observed Earth at several different phases, including near-full and crescent, and these data also provide the highest spectral resolution UV-visible spectra ever published from distant spacecraft observations of our planet. Thus, the *LCROSS* dataset is extremely useful for validating models, as it provides measurements of Earth's appearance at the extremes of phase, with near-simultaneous data across a very wide range of wavelengths.

In addition to their utility for model validation, the *LCROSS* observations also provide a unique opportunity to directly explore Earth's detectable planetary characteristics at wavelengths spanning the UV to infrared. Most significantly, the near-full to crescent phase data allow us to test proposed observational techniques that would constrain the habitability of terrestrial exoplanets by studying the phase- and wavelength-dependent strength of specular reflection (glint) from a planetary ocean. Additionally, the UV-visible portion of the Earth's LCROSS spectra provide a direct measurement of several significant biosignatures, including ozone features, which could be targets for future exoplanet characterization missions (Heap et al. 2008). These data can therefore be used to test characterization and analysis methods for a terrestrial planetary target, Earth, for which we have ground truth data for the surface and atmospheric state.

We describe the *LCROSS* observations in Section 2, and a significant recalibration of the UV-visible spectra in Section 2.1. We briefly describe our model, the Virtual Planetary Laboratory (VPL) three-dimensional spectral Earth model, in Section 3. Model validation is presented in Section 4.1, and we discuss the role of glint in the crescent phase observations in Section 4.2. Section 4.3 describes the detection of ozone absorption in the UV-visible spectra. These findings are discussed in Section 5.

## 2 DESCRIPTION AND RECALIBRATION OF *LCROSS* OBSERVATIONS

The *LCROSS* mission was co-launched with the Lunar Reconnaissance Orbiter (LRO) on June 18, 2009. After launch, *LCROSS* was placed on a "lunar gravity-assist, lunar return orbit" trajectory that positioned the *LCROSS* spacecraft in orbit about the Earth at lunar distance for a period of 108 days. During this time, as part of calibration and pointing campaigns, *LCROSS* observed Earth three times on dates 2009-Aug-1, 2009-Aug-17, and 2009-Sep-18, which will be referred to as Earthlook 1, 2, and 3, respectively. For all three Earthlook sequences, the number and cadence of the images & spectra were tailored to meet the 60 kbps downlink requirement and to test pointing and instrument health. Top-level parameters of the Earth observations are summarized in Table 1.

The *LCROSS* scientific payload consisted of five cameras, three spectrometers, and a high-speed photometer. All of the instruments, except the NSP2 (near-infrared spectrometer #2) were bore-sighted along the +X spacecraft axis (Ennico et al. 2012). The +X direction was the main axis for observing the lunar impact data (nadir), and NSP2 was deliberately aligned in the +Z direction to capture the impact cloud back-illuminated by sunlight. For the Earth calibrations, all five cameras and the two nadir-spectrometers, UV-visible (VSP) and NIR (NSP1), took observations of Earth. Table 2 summarizes the characteristics of the *LCROSS* payload instruments that obtained images or spectra of the Earth. The two nadir spectrometer's 1° fields-of-view (FOV) were co-boresighted. Since the Earth subtended more than 1° during all Earthlook observations, the spectrometers observed only a fraction of the Earth's disk in each exposure. All cameras had large enough FOVs to capture the entirety of Earth's disk.

For Earthlook 1, the majority of the data taken with science potential occurred in the last 39 minutes of the hour long observing sequence and consisted of two series of North-South and East-West slews across Earth's disk, with controlled stares at Earth's center. The two nadir spectrometers took data throughout the crossing patterns, but the five cameras took "snapshots" only at the ends of each limb crossing. The Earth appeared nearly full, at phase angle 23°, with a diameter of 2.2°, subtending 48 pixels in the visible and NIR cameras and 22 pixels in the MIR cameras. During the scans, the NSP1instrument took spectra every 2.35 seconds with a fixed integration time of 588 ms. The visible spectrometer took observing triplet exposures of 100 ms, 500 ms and 2500 ms, repeated every 7 seconds. All but the 100 ms spectra were fully or partially saturated across the wavelength range.

In Earthlook 2 both Earth and the Moon were visible in the wider FOV of the cameras, with the Earth and Moon centers separated by 4.8°. Imaging observations were taken during two 10-minute stare periods centered on Earth and the Moon. For the visible cameras, the Earth (Moon) spanned 35 (3) pixels, and 15 (1) pixels for the MIR cameras. The NSP1 instrument took spectra every 5 seconds during the 10 minute Moon stare, subsequent slew to Earth, and 10 minute Earth stare. Of the 185 scientifically useful spectra only six included a clear view of Earth's illuminated crescent. The visible spectrometer took successive 100/500/2500 ms observing triplets, but now repeated every 10 seconds. Data from nearly all exposures were unsaturated, but only 14 triplets were obtained with the illuminated Earth crescent either fully or partially in the FOV.

For Earthlook 3, Earth was observed with a phase angle of 75 degrees (gibbous) with a total data acquisition period of ~40 minutes on the Earth. The same observing parameters used in Earthlook 2 were used for the spectrometers, with NSP1 acquiring spectra every 5 seconds. The visible spectrometer acquired 100/500/1500 ms exposure triplets every 10 seconds. The camera images resolved Earth with 32 pixels across its diameter for the NIR and visible cameras, and 14 pixels for the MIR cameras.

As described above, in all Earthlooks not all data taken was scientifically usable, due to saturation, underexposure, test patterns, or the Earth not being in the FOV. A summary of the available scientifically usable data for each Earthlook and each instrument is given in Table 3.

Figure 1 is a sample subset of images from Earthlook 1, 2, and 3 in which the Earth spanned 2.2, 1.6, and 1.5 degrees, respectively. Certain regions that appear bright in reflected light seen in the NIR images then appear dark in the emitted radiation seen in the MIR images. Note that, since Earth is emitting radiation at the wavelengths spanned by the MIR cameras, we see the entire disk of the planet (sunlit and un-illuminated portions) in the MIR images.

## *2.1 Recalibration of Visible Spectra*

Initial comparisons between the Earth model and the *LCROSS* UV-visible Earth spectra showed a significant wavelength-dependent discrepancy. Subsequent comparisons between *LCROSS* lunar observations and an independent model of reflectance by the Moon's surface revealed a similar discrepancy, thus indicating that there was a wavelength-dependent offset present in the *LCROSS* UV-visible spectra. This offset was significant enough to frustrate attempts at performing meaningful comparisons between Earth observations and models. The cause of the discrepancy may be related to the visible spectrometer radiance calibration procedure, which was based on an engineering test unit (Ennico et al. 2012), not the true flight unit. To correct the offset, we use a recently published empirical model of the Moon's reflectance (Buratti et al. 2011), along with data taken in flight during a so-called lunar "swingby," where we know the viewing geometry, to re-calibrate the *LCROSS* VSP observations.

High signal-to-noise observations were obtained for the Mendeleev Crater (Swingby 1), Goddard Crater (Swingby 2), and Giordano Bruno Crater (Swingby 3) regions of the Moon on 2009-Jun-23. Table 4 shows the sub-spacecraft, boresight, and sub-solar latitudes and longitudes for the three swingby datasets. As an example, Figure 2 shows a VSP spectrum from Swingby 1 (prior to recalibration), and a corresponding model of the spectrum of the lunar surface. The difference between the data and model is largest near 0.37 µm (370 nm), with the data being too dim by a factor of ~2.

Figure 3 shows wavelength-dependent ratios of the observed spectra to the modeled spectra for all three lunar swingby datasets. Smoothed and un-smoothed curves are shown for Swingby 1, while only smoothed curves are shown for Swingby 2 and 3. Fine structure in the un-smoothed curve is primarily due to noise and differences between the solar spectrum at the time of observation and the spectrum used in our model (Wehrli 1985) and/or small offsets in the wavelength calibration of the instrument. This structure contains little information regarding intensity calibration errors. Thus, we smoothed over the fine structure for each swingby. Smoothed curves for each swingby observation are typically within 5% of their collective mean, so averaging these together yields a wavelength-dependent scaling factor that is both smoothly varying and suitable for re-calibrating *LCROSS* VSP spectra. All visible spectra shown in later portions of this paper have been scaled using this wavelength-dependent factor (made available at vpl.astro.washington.edu).

The primary result to come from *LCROSS* visible spectroscopic dataset is an estimate of the amount of water in the lunar regolith.  Analysis of an OH emission feature near 0.31 μm (310 nm) in the ejecta plume 180 s after the Centaur imact yielded an estimate of 23 +/- 11 kg of water in the instrument field-of-view (Colaprete et al. 2010).  Using our updated calibration, this estimate increases to 29 +/- 12 kg, which is in better agreement with the value derived from *LCROSS* NIR observations at the same time post impact (52.5 +/- 2.6 kg) (T. Colaprete, personal communication).

## 3 THE VPL EARTH MODEL

We use the VPL three-dimensional spectral Earth model to simulate the *LCROSS* Earth observations.  This model is described in detail in Robinson et al. (2011), so only a brief summary of the model is presented below.  Recent model upgrades are discussed in Section 3.2.

### *3.1 Model Description*

The VPL Earth model is a comprehensive tool for simulating the time-dependent appearance of the distant Earth.  The model incorporates absorption and scattering by the atmosphere, clouds, and surface, and includes specular reflectance from the ocean and direction-dependent scattering by clouds, both of which are crucial for determining Earth's phase-dependent brightness in reflected light.  The VPL Earth model has been previously validated against several datasets: visible photometric and NIR spectroscopic observations of Earth from NASA's EPOXI mission, mid-infrared spectroscopic observations of Earth from the Atmospheric Infrared Sounder (AIRS) aboard NASA's *Aqua* satellite (Robinson et al. 2011), and broadband visible observations of Earth's brightness from Earthshine observations (Robinson et al. 2010).

Typically, the VPL Earth model is used to generate phase-dependent disk-integrated spectra of the Pale Blue Dot. To create these spectra, the model computes the integral of the projected area weighted intensity in the direction of the observer, which can be written as

$$F_\lambda(\hat{\mathbf{o}},\hat{\mathbf{s}}) = \frac{R_e^2}{d^2} \int_{2\pi} I_\lambda(\hat{\mathbf{n}},\hat{\mathbf{o}},\hat{\mathbf{s}})(\hat{\mathbf{n}}\cdot\hat{\mathbf{o}})d\omega \qquad (1)$$

where $F_\lambda$ is the disk-integrated specific flux density ("flux" hereafter), $R_e$ is the radius of Earth (6378 km), $d$ is the distance from the observer to the planet, $I_\lambda(\hat{\mathbf{n}},\hat{\mathbf{o}},\hat{\mathbf{s}})$ is the location-dependent specific intensity in the direction of the observer, $d\omega$ is an infinitesimally small unit of solid angle on the globe, $\hat{\mathbf{n}}$ is a surface normal unit vector for the portion of the surface corresponding to $d\omega$, and $\hat{\mathbf{o}}$ and $\hat{\mathbf{s}}$ are unit vectors in the direction of the observer and the Sun, respectively. For a disk-integrated spectrum, the integral in Equation (1) is over the entire observable hemisphere (2π steradians, as indicated), although spectra of smaller regions of the planet's disk can be obtained by integrating over a smaller solid angle (see Section 3.2). Note that the dot product at the end of Equation (1) ensures that an element of area $R_e^2 d\omega$ near the limb is weighted less than an element of equal size near the sub-observer point.

The integral in Equation (1) is numerically evaluated by dividing the globe into sets of equal-area pixels according to the Hierarchical Equal Area isoLatitude Pixelization (HEALPix) model (Górski et al. 2005)[1]. The Earth model maintains two different pixelizations, or resolutions—a set of atmospheric pixels and a set of surface pixels, which are nested beneath the atmospheric pixels. Key atmospheric properties (e.g., temperature and gas mixing ratio profiles) are specified by data from several different Earth-observing satellites, and are mapped at the resolution of the atmospheric pixels (typically 48 pixels over the globe). Coverages of different surface and cloud types (e.g., snow, thin ice clouds) are also taken from Earth-observing satellite data, but are mapped at the higher resolution of the surface pixels. Using this nested pixelization scheme allows us to capture a large dynamic range in surface coverage and cloudiness (which vary on relatively small spatial scales), while still maintaining short model runtimes (which depend primarily on the atmospheric resolution).

Dividing the planet into *N* equal-area surface pixels converts the integral in Equation (1) into a sum over the surface pixels, which is given by

$$F_\lambda(\hat{\mathbf{o}},\hat{\mathbf{s}}) = \frac{4\pi}{N} \frac{R_e^2}{d^2} \sum_{i \in O} I_\lambda(\hat{\mathbf{n}}_i,\hat{\mathbf{o}},\hat{\mathbf{s}})(\hat{\mathbf{n}}_i \cdot \hat{\mathbf{o}}) \; , \qquad (2)$$

where $\hat{\mathbf{n}}_i$ describes the location on the sphere of surface pixel *i*, and *O* is the set of indices of all observable pixels (i.e., all pixels with $\hat{\mathbf{n}}_i \cdot \hat{\mathbf{o}} > 0$ ). The leading factor in this expression comes from the solid angle of each equal-area pixel, which is equal to $4\pi/N$.

For each modeled date, we generate a database of synthetic surface pixel spectra using the Spectral Mapping Atmospheric Radiative Transfer (SMART) model (developed by D. Crisp, see Meadows & Crisp 1996), which is a one-dimensional, fully multiple-scattering, line-by-line radiative transfer model. For each atmospheric pixel, we run the SMART model over a grid of surface types, cloud types, observer and solar zenith angles, and azimuth angles. The spectrum of a surface pixel (i.e, $I_\lambda(\hat{\mathbf{n}}_i,\hat{\mathbf{o}},\hat{\mathbf{s}})$ in Equation (2)) beneath this atmospheric pixel is then obtained by interpolating over the spectral database to the particular viewing geometry for the pixel, and by weighting interpolated spectra according to the coverage of different surface and cloud types on that surface pixel.

---

[1] see http://healpix.sourceforge.net

### 3.2 Model Upgrades

We have made two substantial upgrades to the VPL Earth model published in Robinson et al. (2011). First, we now use the HITRAN 2008 line list database (Rothman et al. 2009), whereas previous versions of our model used HITRAN 2004. Second, we have upgraded the model, which previously provided only full disk-integrated spectra as output, to allow for integration of pixel intensities over a smaller, user-specified region of Earth's disk. This enables comparisons between the model and the partial-disk *LCROSS* observations. In this case, the integral in Equation (1) is not over 2π steradians, but is instead over the smaller solid angle encompassed by a user-defined FOV. For the *LCROSS* spacecraft, this is a 1° diameter circular aperture.

To produce partial-disk observations, we use the Earth model to generate a datacube of pixel spectra for a user-defined observation time and viewing geometry. This datacube contains all of the spectra of the individual surface pixels as seen from the specified vantage point (i.e., $I_\lambda(\hat{\mathbf{n}},\hat{\mathbf{o}},\hat{\mathbf{s}})$ for each pixel). Using the HEALPix model, we determine which surface pixels are contained within a specified FOV, and then combine the solid angle weighted spectra of these pixels. Figure 4 demonstrates this process, and shows a whole-disk view of Earth as seen from the *LCROSS* spacecraft during the first epoch of observation (Earthlook 1), as well as a partial-disk view corresponding to the spectrometer FOV. The associated whole-disk and partial-disk spectra are also shown.

## 4 RESULTS

### 4.1 *Images and Model Validation with Earthlook 1 and 3*

Model images of Earth corresponding to the NIR2 and MIR1 cameras, as well as simulated true color images, are shown for the three Earthlooks in Figure 5. Note that the model images are at higher spatial resolution than the data images, which were presented in Figure 1. Unfortunately, issues with absolute radiometric calibration prevent meaningful quantitative comparisons with the models (see technical report by Ennico et al. 2010)[2]. Instead, we use the *LCROSS* camera images as they were intended—to provide context.

Images from Earthlook 1, where Earth is viewed at near-full phase, show primarily a view of the Pacific Ocean, with land in North America and the northern polar cap also present. In the NIR, model images reveal that brightness is dominated by liquid water cloud structures off the western coast of North America and over northeastern Asia. Smaller structures of ice clouds are present in the equatorial Pacific, which are quite distinct in the model MIR images, where clouds appear as dark features due to their location at colder altitudes. Note that, at visible and shorter NIR wavelengths, droplets in liquid water clouds tend to have single-scattering albedos near unity, and also tend to forward scatter. Crystals in ice clouds behave similarly, except for a distinct decrease in their single-scattering albedos near 1.6 µm. Both particle types have markedly lower single-scattering albedos (~0.2–0.5) in the MIR. For a review of the optical properties of Earth's clouds, see Kokhanovsky (2004).

---

[2] http://ntrs.nasa.gov/archive/nasa/casi.ntrs.nasa.gov/20110008670_2011009183.pdf

The view in Earthlook 3 is of South America and the northern Atlantic Ocean, and Earth is viewed near quadrature. The brightest features in the NIR are liquid water clouds over South America. A few ice cloud structures appear near Central America. In Earthlook 2, which is a crescent phase observation, specular reflectance from the Indian Ocean (glint) dominates the NIR data image and is reproduced in the model. Glint, which originates at the surface, is relatively easy to detect using the NIR2 camera's wide bandpass since the effect is most pronounced at wavelengths that are unaffected by Rayleigh scattering and strong gas absorption.

While the large-scale pixelization in the simulated MIR images is due to spatial averaging of atmospheric properties (which is necessary to reduce model runtime), a distinct meridional gradient can be seen. This gradient is also quite apparent in the *LCROSS* images for Earthlook 2. The northern hemisphere is relatively warm, as it is transitioning from summer to fall, while the southern hemisphere is relatively cool, as it is transitioning from winter to spring. Thus, this gradient, which influences MIR observations, is linked to seasonality, and will vary in time (Cowan et al. 2012b).

We performed a quantitative validation of the Earth model against the observed UV-visible and NIR spectra for Earthlook 1 and 3. (Comparisons to Earthlook 2 are given special treatment due to the importance of glint, and are presented in the following section.) Figure 6 shows the spectrometer FOV and the position of the spectrometer boresight center through time for each Earthlook. Ultraviolet-visible and NIR spectra are shown for Earthlook 1 and Earthlook 3 in Figure 7 and Figure 8, respectively. The shade of each spectrum corresponds to the pointings shown in Figure 6.

Spacecraft pointing was being tested during Earthlook 1, so the boresight wandered over a large portion of Earth's disk. As a result, we see a larger dynamic range in Earthlook 1 than in Earthlook 3, where the pointing was more tightly constrained. In Earthlook 1, cloud features dominate the brightest spectra, while the dimmest spectra are for pointings near Earth's limb. For Earthlook 3, the brightest spectra correspond to pointings that contain a large cloud feature on the central-western coast of South America that is close to the edge of the FOV.

The most notable absorption feature in the UV-visible spectra is the 0.255 μm (255 nm) ozone feature, with the remainder of the features being primarily due to the solar spectrum. The NIR spectra clearly show water vapor absorption bands at 1.4 μm and 1.9 μm. Signal-to-noise ratios (SNRs) are largest outside of these bands, especially in the continuum region near 1.6 μm, and throughout most of the visible range.

Model spectra for Earthlook 1 and 3 are also shown in Figure 7 and Figure 8. As there are several hundred observed spectra for each of these Earthlooks, we chose to limit our comparison to the extremes in brightness, thus demonstrating that the model captures the absolute brightness and the dynamic range. The boresight center latitude and longitude for the brightest and dimmest spectra (with regard to continuum) were used to generate model spectra, integrated over the same sub-region of Earth's disk as the observations.

Overall, the model spectra closely resemble the observations, and capture the dynamic range seen in Earthlook 1 and 3. The model falls within the 10% calibration uncertainty in the data in most cases. One exception is the dimmest model spectrum for Earthlook 1, which is about 30% too dim in the NIR continuum region near 1.6 µm (Figure 7). This discrepancy is most likely due to pointing uncertainties, which are 5–10% the size of the FOV diameter (1°). As the dimmest pointings for Earthlook 1 occur when the spectrometer is at the limb, where only about half of the FOV covers Earth's disk, small pointing errors translate to relatively large changes in brightness, which we confirmed with the model.

### 4.2    Glint

Hundreds of spectral observations were taken during Earthlook 2, and most were centered on un-illuminated regions of the disk. Only a small number of these observations clearly captured the illuminated crescent of Earth—one UV-visible spectrum and six NIR spectra. These spectra are shown in Figure 9, along with model spectra. The standard VPL Earth model includes specular reflectance from the ocean surface, or glint, and we also show a model where the glint effect has been removed. In the model without glint, the ocean surface is assumed to scatter in an isotropic (Lambertian) fashion, using a measured wavelength-dependent ocean albedo.

At most continuum wavelengths, only the model that includes glint can match the observations. This model is within the 10% calibration uncertainty for the observations, and matches the UV-visible spectrum especially well. Both models reproduce the observations shortward of 0.4 μm (400 nm) and in the NIR water absorption bands. At these wavelengths, the observations are not sensitive to the surface, where the glint effect originates. This lack of sensitivity to the surface is due to Rayleigh scattering and ozone absorption, which dominate the spectrum shortward of 0.4 μm, and the optically-thick NIR water bands. Thus, it would be expected that a model without glint would reproduce crescent-phase observations of Earth at these wavelengths. Longward of 0.4 μm, however, the inclusion of glint is essential for reproducing the observations at continuum wavelengths. For the visible spectra, the glint effect increases the brightness by as much as 40%, with the contribution growing to as much as 80% in the NIR spectra. This relative increase also holds for the disk-integrated spectra (shown in the NIR panel of Figure 9), although the overall brightness decreases due to integration over the unilluminated portion of the disk.

### 4.3 Ozone Detection

The most striking feature in the observed UV-visible spectra is the 255 nm near-UV ozone absorption feature. To better demonstrate this feature, we show reflectance spectra from all three Earthlooks in Figure 10. These spectra were generated by first selecting intermediate brightness cases from each Earthlook. Dividing these spectra by the lunar spectrum from Swingby 1 removes the solar spectrum, and produces a quantity that is proportional to Earth's reflectivity divided by the Moon's reflectivity. We remove the contribution from the lunar reflectance by using the previously-mentioned reflectivity model from Buratti et al. (2011). While a small amount of noise remains, the near-UV ozone feature is clearly visible, as well as two water vapor bands and an oxygen ($O_2$) feature.

The *LCROSS* observations are dominated by systematic noise in the near-UV ozone band, while the model shows that Earth's brightness falls by about 4–5 orders of magnitude between 450 nm and the core of the Hartley band. This decrease in brightness is due primarily to ozone absorption, but the solar spectrum is also decreasing at these wavelengths. As a result, the reflectivity decrease, which divides out the solar flux, between these two wavelengths, while still very dramatic, is slightly less exaggerated. Roughly $10^{17}$ molecules $cm^{-2}$ are required to make the Hartley ozone band optically thick, compared to typical levels of about $10^{19}$ molecules $cm^{-2}$ for Earth's atmosphere. Stratospheric ozone is produced by photodissociation of oxygen molecules and a series of chemical reactions leading to the combination of an oxygen atom with an oxygen molecule, yielding ozone (Solomon 1999). The kinetics of this reaction in Earth's atmosphere is such that molecular oxygen at a level of only ~0.1% that of the present atmospheric level (which is 21% by volume) would be required to produce ozone concentrations high enough to make the Hartley band optically thick (Kasting et al. 1985; Kaltenegger et al. 2007) and, hence, more easily detected.

Figure 10 also shows model reflectance spectra for Earthlook 3 both with and without ozone absorption included. The Hartley band is clearly the dominant feature in the near-UV. Comparing the models with and without ozone shows the presence of the Chappuis band (400–700 nm). This absorption feature is both broad and weak, and influences the spectrum at levels less than about 10%. As this difference is of the order of the data and model uncertainties, it is difficult to detect the Chappuis band in the observations.

## 5 DISCUSSION

### *5.1 Data-Model Comparisons*

The VPL Earth model has been previously validated against a number of datasets. In Robinson et al. (2011), the model was compared broadband photometric UV and visible observations of Earth. The spectral resolution of the UV-visible spectra presented in this work is significantly larger than the *EPOXI* observations ($\lambda/\Delta\lambda \sim 500$ vs. 6), and the good agreement between the *LCROSS* data and model at UV and visible wavelengths lends further credibility to the model. In the *EPOXI* comparisons, the model was systematically dimmer than the observations at 350 nm and 450 nm by about 6–8%, and Fujii et al. (2011) found similar offsets at these wavelengths. However, our model agrees with the *LCROSS* spectra at these wavelengths. We therefore conclude that the *EPOXI* data may contain a small calibration offset in absolute brightness at UV wavelengths. This conclusion is somewhat weak, though, since both the *EPOXI* and *LCROSS* observations are only absolutely calibrated to the ~10% level.

Comparisons of the model to high spectral resolution observations of the glinting Earth at crescent phase are the first of their kind. We find very good agreement between the data and model in the crescent phase data set (Earthlook 2), especially at visible wavelengths, where discrepancies are well within the 10% calibration uncertainty. This comparison confirms model predictions of Earth's brightness made in Robinson et al. (2010). We note that, at the phase angle of Earthlook 2 (129°), observing an Earth-twin in an edge-on orbit around a Sun-like star at 10 pc would require a telescope inner working angle of at least 77 mas. This estimate doubles if the system is instead taken to be at 5 pc. These inner working angles are comparable to those being considered for future terrestrial exoplanet characterization mission concepts (e.g., Brown & Soummer 2010), and so searching for glint at this phase is theoretically possible for such missions.

Previous model validations against broadband (400–700 nm) Earthshine observations showed that the VPL Earth model was systematically 20–35% dimmer than the data at crescent phases. The agreement we find here between the crescent phase *LCROSS* observations and our Earth model suggests that the systematic offset between the model and the Earthshine observations may be due to a calibration issue in the Earthshine data, which could have ramifications for published estimates of Earth's spherical albedo at visible wavelengths (Pallé et al. 2003). We estimate that these systematics could result in errors in the spherical albedo measurement as large as 0.01–0.02 (or about 3–6%). While not a large effect overall, these errors are much larger than the reported uncertainties, which were typically ~1%. Further space-based observations of the crescent phase Earth at visible wavelengths would help to clarify this issue.

The crescent phase data-model comparisons confirm the predictions of Earth's brightness made in Robinson et al. (2010), and demonstrate the significant contribution of glint to Earth's crescent phase reflectivity. In the context of exoplanet habitability, there are, however, effects that could be false positives for the glint signal, such as cloud scattering (Robinson et al. 2010) and the tendency of crescent phase observations to probe colder regions of a planet (Cowan et al. 2012a). Thus, it remains to be shown whether or not glint can be used to unambiguously detect oceans on exoplanets. Approaches to disentangling potential false positives from the glint effect may include using polarization signatures caused by specular reflection (Williams & Gaidos 2008; Zugger et al. 2010). However, it is important to note that both Rayleigh and cloud scattering can have strong polarization signatures, and either of which can overwhelm the glint polarization signature, thus affecting the interpretation of observations (Zugger et al. 2011). Another approach might utilize wavelength-dependence in the glint signature (Robinson et al. 2010), which is likely distinct from cloud or snow/ice reflection.

Satellite images of Earth have revealed that cirrus clouds can also produce specular reflection, which occurs when the ice crystals in the cloud are oriented preferentially in the horizontal direction (Chepfer et al. 1999). While this phenomenon is important to remote sensing in the Earth sciences, it does not significantly influence disk-integrated observations due to the spatial scales involved. As Figure 5 shows, the ocean glint spot appears large at crescent phases, allowing it to strongly influence the crescent Earth's reflectivity. Cirrus clouds, however, are typically ~10 km in horizontal scale (although larger cirrus clouds are known to occur) (Dowling & Radke 1990), and cover <10% of the planet (for moderate and thick cirrus clouds) (Chen et al. 2000). Furthermore, it would seem extremely unlikely that preferential horizontal orientation of ice crystals could be maintained across the large number of cirrus structures required to cover an area similar to that as the ocean glint spot at crescent phases.

### 5.2 The Ozone Hartley Band as a Biosignature

The 255 nm Hartley absorption feature due to ozone in the Earth's atmosphere is noteworthy for its strength. If seen in the atmosphere of an exoplanet, this band would provide evidence for life, since, at least on Earth, it is produced via the photolysis of photosynthetically produced oxygen. Heap et al. (2008) argue that the ozone Hartley band may be the most sensitive tracer of oxygen in an exoplanet's atmosphere, and it is certainly one of the strongest absorption features in Earth's spectrum. There are also a number of observational advantages to searching for UV features on extrasolar planets. Since spatial resolution is proportional to wavelength, the UV offers higher spatial resolution for a given telescope diameter and, hence, improved ability to spatially resolve the planet from its host star. There is also a reduced influence of extrazodiacal dust, which has a lower reflectivity at UV wavelengths.

However, there are a number of challenges to observing such a feature in the spectrum of an exoplanet. Stars with temperature at or below that of the Sun emit little flux at UV wavelengths, and UV detectors typically have lower efficiency at these wavelengths than at visible wavelengths. Additionally, many common atmospheric gases strongly absorb UV radiation in the 100–300 nm range, potentially confusing efforts to identify and quantify ozone levels. Because of these factors, spectra with relatively high S/N and spectral resolution are needed to measure the planet's spectrum at these wavelengths with the precision needed to quantify the ozone abundance from such observations.

This possibility of "false positives" is also significant and needs to be further explored to best determine which observational capabilities will be able to discriminate features due to true biologically-mediated ozone production from those due to strong abiotic sources (Domagal-Goldman et al., submitted) or other molecules. For example, in Figure 11 we show that transmission through a column of ozone and a column of sulfur dioxide ($SO_2$), both with a column density of $8 \times 10^{18}$ molecules $cm^{-2}$. The fall-off in transmission near 300nm is nearly identical and would be difficult to discriminate with photometry alone or with very-low resolution spectroscopy. In this case, either more capable instrumentation would be required, or examination of other regions of the planet's spectrum could help to determine the source of the UV feature. For example, as shown in the *LCROSS* spectra, both the UV Hartley band and the Chappuis band (400–700 nm) of ozone are potentially visible for Earth-like quantities of ozone. Additionally, $O_2$ strongly absorbs at wavelengths shortward of 260 nm, and so we would not expect to see a "continuum" on the short wavelength side of the $O_3$ Hartley band if abundant oxygen was also present. However, many other molecules absorb shortward of 200 nm, and while detection of a continuum at wavelengths shortward of this wavelength would likely rule out abundant oxygen, failure to detect the continuum would not necessarily confirm that $O_2$ was present, or that the $O_3$ was biologically mediated.

## 6  CONCLUSIONS

We presented NIR and MIR images as well as UV-visible and NIR spectroscopic observations of Earth taken by the *LCROSS* spacecraft on three separate occasions (2009-Aug-1, 2009-Aug-17, and 2009-Sep-18). These observations span a range of Earth phase from crescent to near-full. A significant recalibration was performed for the UV-visible spectra to correct a wavelength dependent offset in the original calibration. The *LCROSS* Earth dataset was then used to perform the first high-spectral resolution, visible wavelength validation of the VPL 3-D spectral Earth model. The model was shown to reproduce the absolute brightness and dynamic range of the spectroscopic observations from all epochs of observation to within the ~10% data absolute calibration uncertainty. Data-model comparisons using the crescent phase observations revealed strong specular reflectance from Earth's oceans. This confirms wavelength-dependent model predictions at these phases, and provides an observational test of a technique that might be used to detect surface oceans on exoplanets. The validated model is now a dependable stand-in for data in studies that aim to understand planetary properties from observations, such as a mapping of cloud and surface features (e.g., Kawahara & Fujii 2011) or the retrieval of the atmospheric state (Madhusudhan & Seager 2009; Lee et al. 2012; Line et al. 2012; von Paris et al. 2013). We discussed the detection and strength of the 255 nm ozone Hartley band in Earth's spectrum as an analog for detection of a photosynthetic biosphere on an extrasolar planet. We conclude that detection of such a feature would be intriguing, and would certainly indicate that a more detailed observational follow-up of the planet was warranted. However, it would not be a definitive detection of an oxygenic photosynthetic biosphere unless the spectral band could be measured and the $O_3$ abundance quantified, and/or other corroborating features of the planetary spectra were observed to support a likely biologically mediated source for the observed ozone.

## ACKNOWLEDGEMENTS

This work was performed as part of the NASA Astrobiology Institute's Virtual Planetary Laboratory, supported by the National Aeronautics and Space Administration through the NASA Astrobiology Institute under solicitation No. NNH05ZDA001C. This research was also supported by the NASA Lunar Science Institute (NLSI) under solicitation No. NNH08ZDA008C, "Scientific and Exploration Potential of the Lunar Poles" (P.I. B. Bussey). TR gratefully acknowledges support from an appointment to the NASA Postdoctoral Program at the Ames Research Center, administered by Oak Ridge Associated Universities. STScI is operated by the Association for Universities of Research in Astronomy, Inc., under NASA contract NAS5-26555. Some of the results in this paper have been derived using the HEALPix (Gorski et al., 2005, ApJ, 622, p759) package.

## REFERENCES

Arnold, L., Gillet, S., Lardiere, O., Riaud, P., & Schneider, J. 2002, Astronomy and Astrophysics, 392, 231
Basri, G., Borucki, W. J., & Koch, D. 2005, New Astronomy Review, 49, 478
Batalha, N. M., et al. 2013, The Astrophysical Journal Supplement Series, 204, 24
Borucki, W. J., et al. 2010, Science, 327, 977
Borucki, W. J., et al. 2011a, The Astrophysical Journal, 728, 117
Borucki, W. J., et al. 2011b, The Astrophysical Journal, 736, 19
Borucki, W. J., et al. 2003, in Society of Photo-Optical Instrumentation Engineers (SPIE) Conference Series, eds. J. C. Blades, & O. H. W. Siegmund, 129
Brown, R. A., & Soummer, R. 2010, The Astrophysical Journal, 715, 122

Buratti, B. J., et al. 2011, Journal of Geophysical Research (Planets), 116, E00G03
Chen, T., Rossow, W. B., & Zhang, Y. 2000, Journal of Climate, 13, 264
Chepfer, H., Brogniez, G., Goloub, P., Breon, F. M., & Flamant, P. H. 1999, Journal of Quantitative Spectroscopy and Radiative Transfer, 63, 521
Christensen, P. R., & Pearl, J. C. 1997, J Geophys Res, 102, 10875
Colaprete, A., Elphic, R. C., Heldmann, J., & Ennico, K. 2012, Space Science Reviews, 167, 3
Colaprete, A., et al. 2010, Science, 330, 463
Cowan, N. B., Abbot, D. S., & Voigt, A. 2012a, The Astrophysical Journal Letters, 752, L3
Cowan, N. B., et al. 2009, Astrophys J, 700, 915
Cowan, N. B., et al. 2011, Astrophysical Journal, 731
Cowan, N. B., & Strait, T. E. 2013, The Astrophysical Journal Letters, 765, L17
Cowan, N. B., Voigt, A., & Abbot, D. S. 2012b, The Astrophysical Journal, 757, 80
Danjon, A. 1928, Ann Obs Strasbourg, 2, 165
Des Marais, D. J., et al. 2002, Astrobiology, 2, 153
Dowling, D. R., & Radke, L. F. 1990, Journal of Applied Meteorology, 29, 970
Dressing, C. D., & Charbonneau, D. 2013, The Astrophysical Journal, 767, 95
Ennico, K., Colaprete, A., Shirley, M., & Wooden, D. 2010, in NASA Technical Report
Ennico, K., Shirley, M., Colaprete, A., & Osetinsky, L. 2012, Space Science Reviews, 167, 23
Ford, E. B., Seager, S., & Turner, E. L. 2001, Nature, 412, 885
Fujii, Y., Kawahara, H., Suto, Y., Fukuda, S., Nakajima, T., Livengood, T. A., & Turner, E. L. 2011, The Astrophysical Journal, 738, 184
Fujii, Y., Kawahara, H., Suto, Y., Taruya, A., Fukuda, S., Nakajima, T., & Turner, E. L. 2010, Astrophysical Journal, 715, 866
Fujii, Y., Turner, E. L., & Suto, Y. 2013, The Astrophysical Journal, 765, 76
Gómez-Leal, I., Pallé, E., & Selsis, F. 2012, The Astrophysical Journal, 752, 28
Goode, P. R., et al. 2001, Geophy Res Lett, 28, 1671
Górski, K. M., Hivon, E., Banday, A. J., Wandelt, B. D., Hansen, F. K., Reinecke, M., & Bartelmann, M. 2005, The Astrophysical Journal, 622, 759
Heap, S. R., Lindler, D., & Lyon, R. 2008. in Astronomical Telescopes and Instrumentation: Synergies Between Ground and Space, Detecting biomarkers in exoplanetary atmospheres with a Terrestrial Planet Finder (International Society for Optics and Photonics), 70101N
Hearty, T., Song, I., Kim, S., & Tinetti, G. 2009, Astrophysical Journal, 693, 1763
Kaltenegger, L., Traub, W. A., & Jucks, K. W. 2007, The Astrophysical Journal, 658, 598
Kasting, J. F., Holland, H. D., & Pinto, J. P. 1985, J Geophys Res, 90, 10
Kawahara, H., & Fujii, Y. 2011, The Astrophysical Journal Letters, 739, L62
Kokhanovsky, A. 2004, Earth-Science Reviews, 64, 189
Kopparapu, R. K. 2013, The Astrophysical Journal Letters, 767, L8
Lee, J. M., Fletcher, L. N., & Irwin, P. G. 2012, Mon Not R Astron Soc, 420, 170
Line, M. R., Zhang, X., Vasisht, G., Natraj, V., Chen, P., & Yung, Y. L. 2012, The Astrophysical Journal, 749, 93
Livengood, T. A., et al. 2011, Astrobiology, 11, 907
Madhusudhan, N., & Seager, S. 2009, Astrophysical Journal, 707, 24
Manalo-Smith, N., Smith, G. L., Tiwari, S. N., & Staylor, W. F. 1998, J Geophys Res, 103, 19733
Marcq, E., Bertaux, J.-L., Montmessin, F., & Belyaev, D. 2012, Nature Geoscience, 6, 25
Meadows, V. S. 2006, in IAU Colloq 200: Direct Imaging of Exoplanets: Science and Techniques, ed. C. A. F. Vakili, 25

Meadows, V. S., & Crisp, D. 1996, J Geophys Res, 101, 4595
Montañés-Rodríguez, P., Pallé, E., Goode, P. R., & Martín-Torres, F. J. 2006, Astrophysical Journal, 651, 544
Oakley, P. H. H., & Cash, W. 2009, Astrophys J, 700, 1428
Pallé, E., et al. 2003, Journal of Geophysical Research (Atmospheres), 108, 4710
Qiu, J., et al. 2003, J Geophys Res, 108, 4709
Robinson, T. D., Meadows, V. S., & Crisp, D. 2010, Astrophysical Journal Letters, 721, L67
Robinson, T. D., et al. 2011, Astrobiology, 11, 393
Rothman, L. S., et al. 2009, Journal Of Quantitative Spectroscopy & Radiative Transfer, 110, 533
Rugheimer, S., Kaltenegger, L., Zsom, A., Segura, A., & Sasselov, D. 2012, Astrobiology, 13, 251
Sagan, C., Thompson, W. R., Carlson, R., Gurnett, D., & Hord, C. 1993, Nature, 365, 715
Sanromá, E., & Pallé, E. 2011, The Astrophysical Journal, 744, 188
Sanromá, E., Pallé, E., & García-Muñoz, A. 2013, The Astrophysical Journal, 766, 133
Seager, S., Turner, E. L., Schafer, J., & Ford, E. B. 2005, Astrobiology, 5, 372
Solomon, S. 1999, Reviews of Geophysics, 37, 275
Stam, D. M. 2008, Astronomy & Astrophysics, 482, 989
Sterzik, M. F., Bagnulo, S., & Palle, E. 2012, Nature, 483, 64
Tinetti, G., Meadows, V. S., Crisp, D., Fong, W., Fishbein, E., Turnbull, M., & Bibring, J.-P. 2006, Astrobiology, 6, 34
Traub, W. A. 2012, The Astrophysical Journal, 745, 20
Turnbull, M. C., et al. 2006, Astrophysical Journal, 644, 551
von Paris, P., Hedelt, P., Selsis, F., Schreier, F., & Trautmann, T. 2013, Astronomy & Astrophysics, 551, A120
Wehrli, C. 1985, in World Radiation Center (WRC) Publications, No 615
Williams, D. M., & Gaidos, E. 2008, Icarus, 195, 927
Woolf, N. J., Smith, P. S., Traub, W. A., & Jucks, K. W. 2002, Astrophysical Journal, 574, 430
Zugger, M. E., Kasting, J. F., Williams, D. M., Kane, T. J., & Philbrick, C. R. 2010, Astrophysical Journal, 723, 1168
---. 2011, Astrophysical Journal, 739, 12

FIGURES

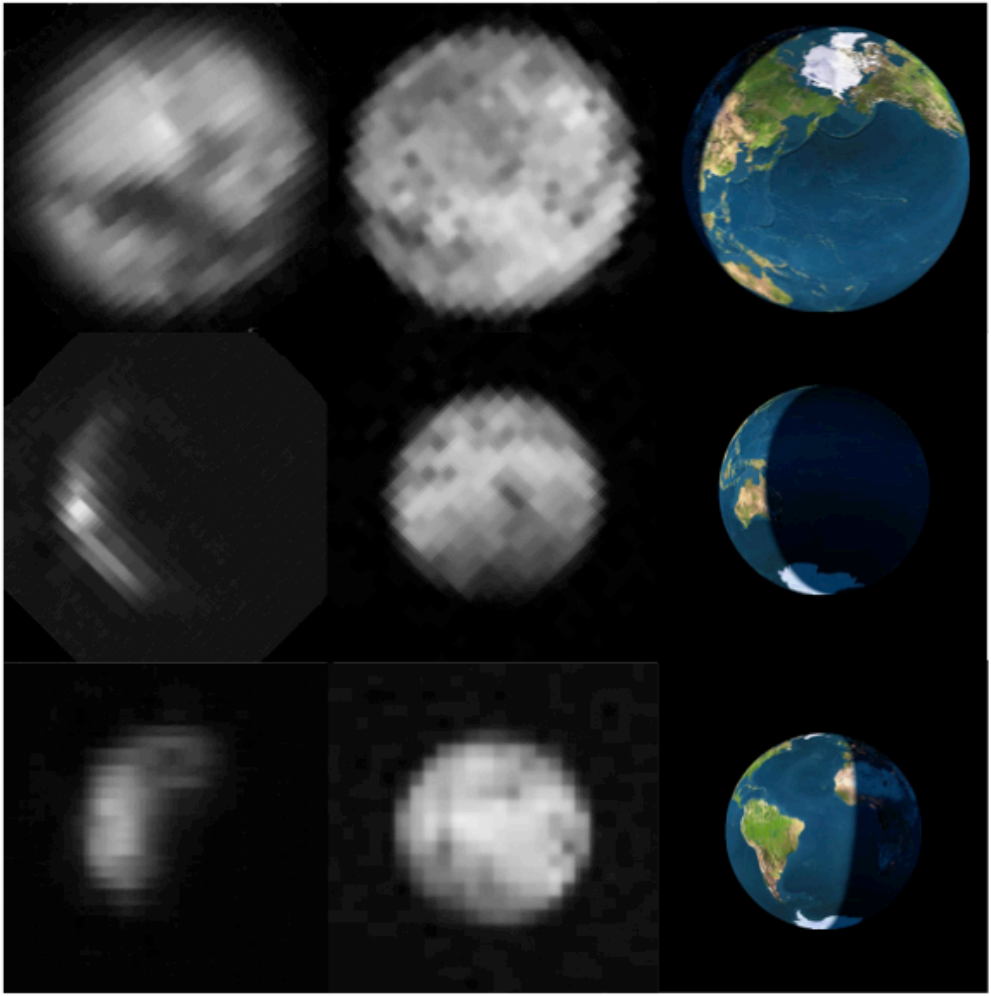

**Figure 1.** Images of Earth taken by NIR2 (0.9–1.7 µm) (left), MIR1 (6.0–10.0 µm) (middle) from each Earthlook. A schematic showing Earth's orientation is also shown (right), which were generated using the Earth and Moon Viewer, first implemented by J. Walker (http://www.fourmilab.ch/cgi-bin/Earth). Rows top to bottom are Earthlook 1 (2009-Aug-1), Earthlook 2 (2009-August-17), and Earthlook 3 (2009- Sep-18) where Earth's angular diameter spanned 2.2, 1.6, and 1.5 degrees, respectively. Each image box represents a 2.5° x2.5° FOV. A glint spot can be seen near the center of the crescent in the Earthlook 2 NIR2 image, and is located in the Indian Ocean off the western coast of Australia.

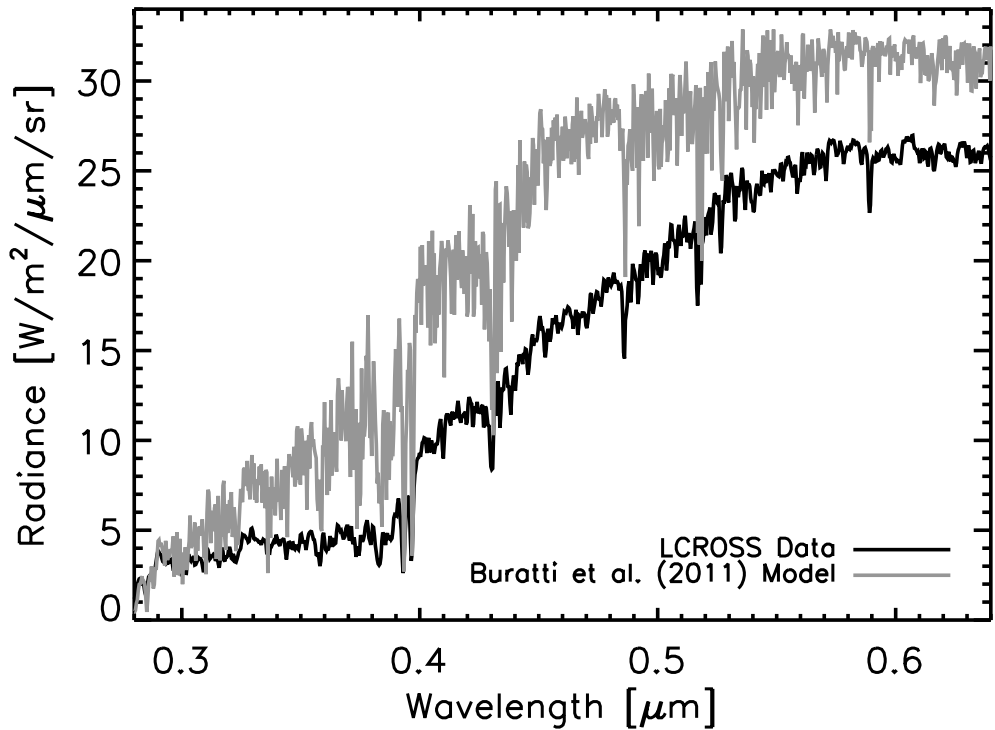

**Figure 2.** Observed (black) and modeled (gray) spectra of the Mendeleev Crater region of the Moon (Swingby 1). We use a validated empirical lunar reflectance model from Buratti et al. (2011). The wavelength-dependent offset between the data and model is due, primarily, to calibration errors in the data.

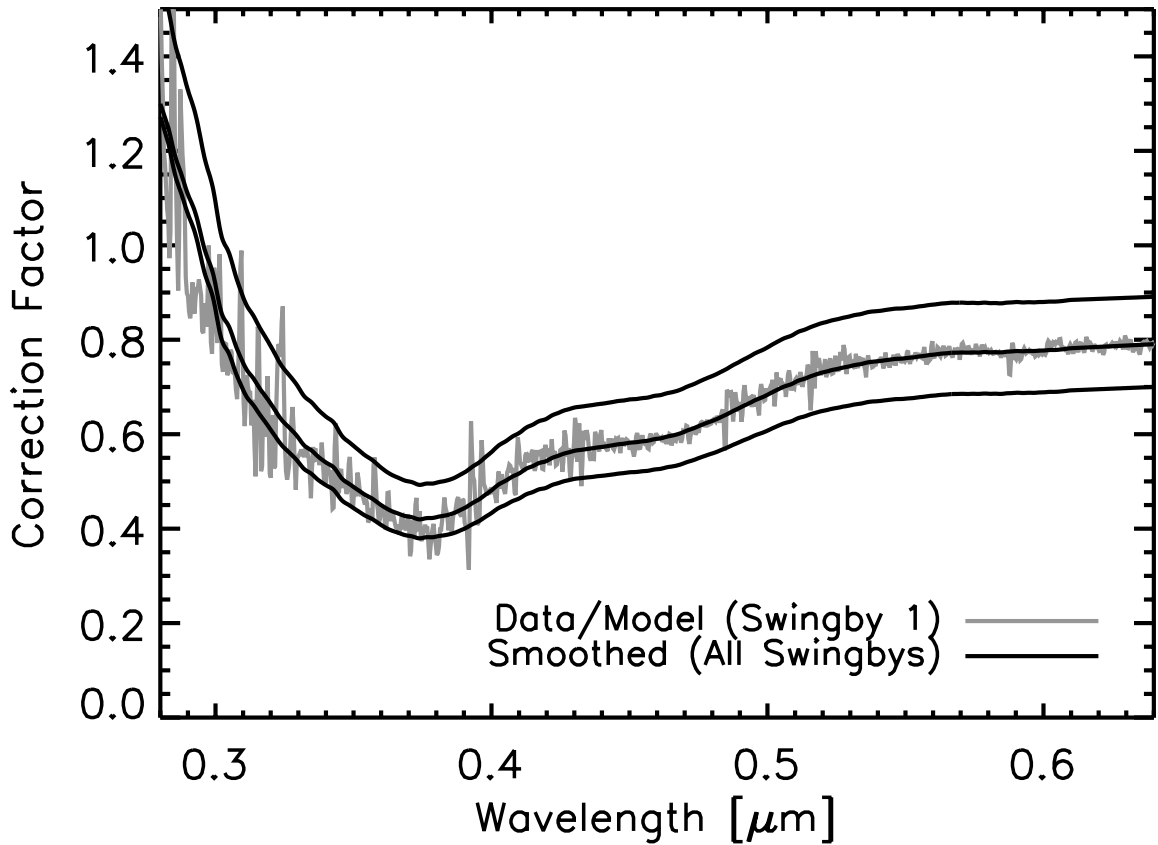

**Figure 3.** Ratio of the data to the model shown in Figure 2 (gray, for Swingby 1) and smoothed ratios (black) for all three swingby datasets. Fine structure in the gray curve is due to primarily to small differences in solar spectra and/or small offsets in the instrument wavelength calibration. The average of the smoothed curves provides a wavelength-dependent scaling that is applied to all later visible spectra.

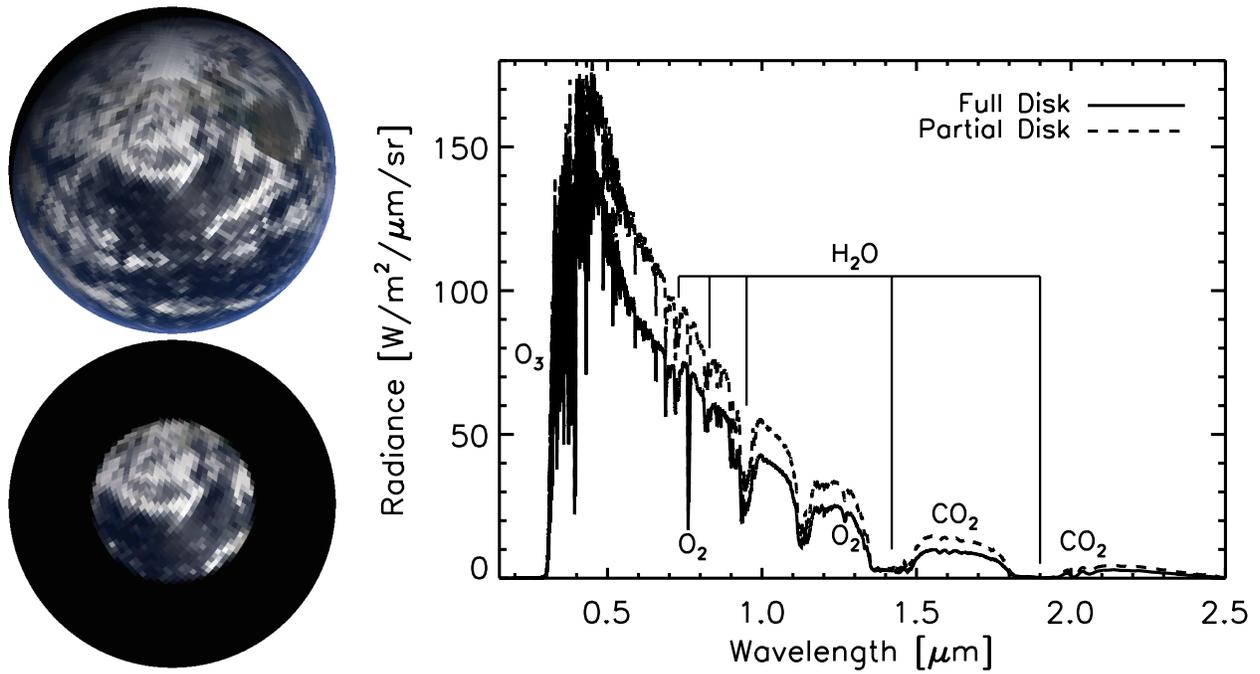

**Figure 4.** Full disk and partial disk views of Earth and corresponding solid angle integrated spectra from the VPL Earth model. The date and viewing geometry correspond to the first *LCROSS* epoch of observation (Earthlook 1). The partial disk view is brighter than the full disk view at most wavelengths as the scene contained within the FOV contains a significant amount of cloud coverage.

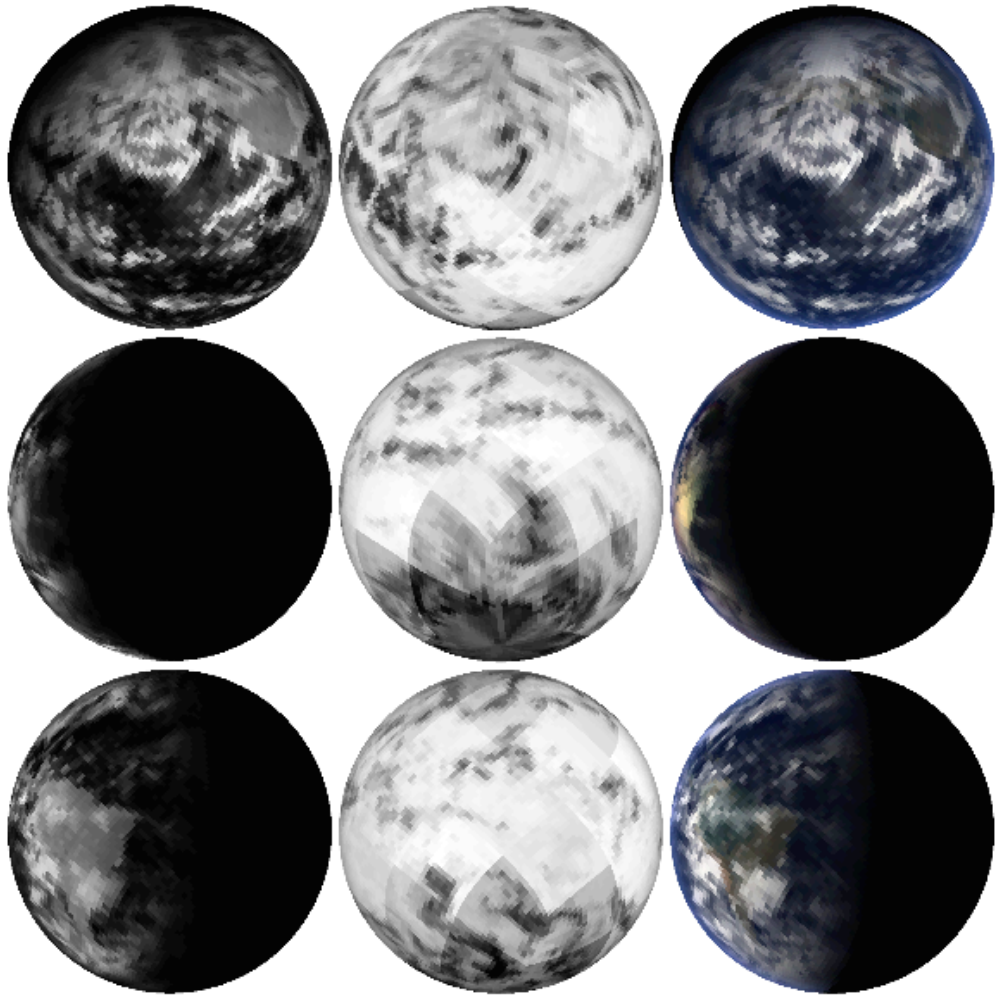

**Figure 5.** Simulated images of Earth from the VPL Earth model for Earthlook 1 (top), Earthlook 2 (middle), and Earthlook 3 (bottom). The image at the right is true color, while the other two images correspond to the NIR2 (0.9–1.7 µm)(left) and MIR1 (6.0–10.0 µm)(middle), as in Figure 1. Note the glint spot in Earthlook 2, near the center of the crescent in NIR and true color images.

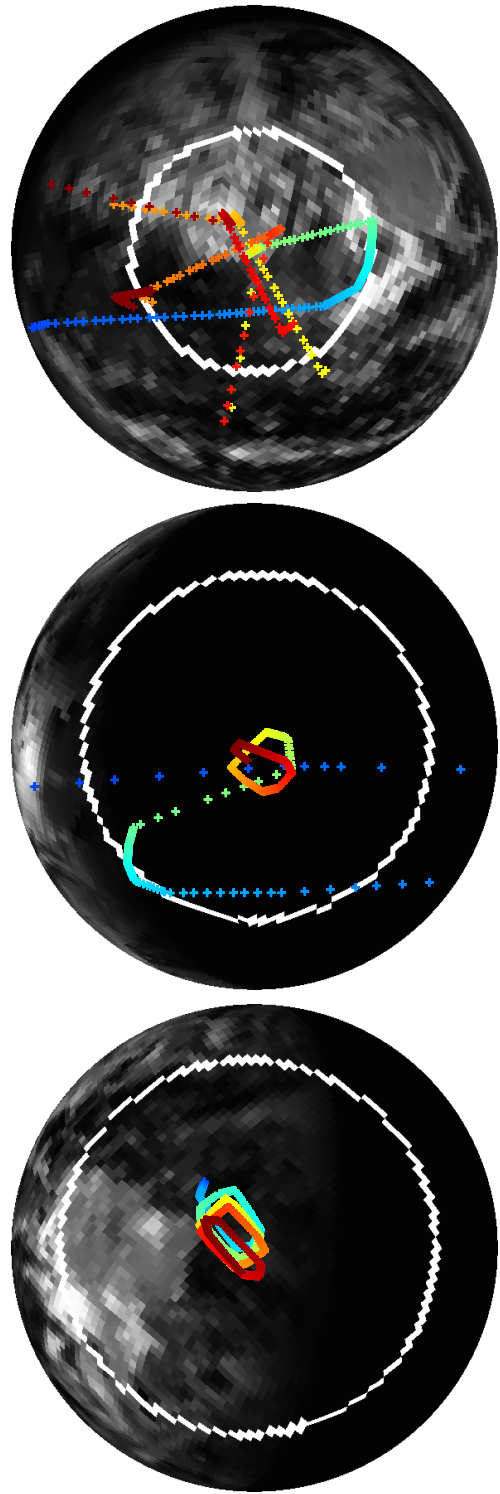

**Figure 6.** Pointing on Earth's disk for Earthlook 1 (top), Earthlook 2 (middle), and Earthlook 3 (bottom). Shades indicate the progression of time during each Earthlook, with darker shades being earlier in the observational sequence. The white ring, which is centered on the disk, indicates the field-of-view of the visible and NIR spectrometers for each Earthlook.

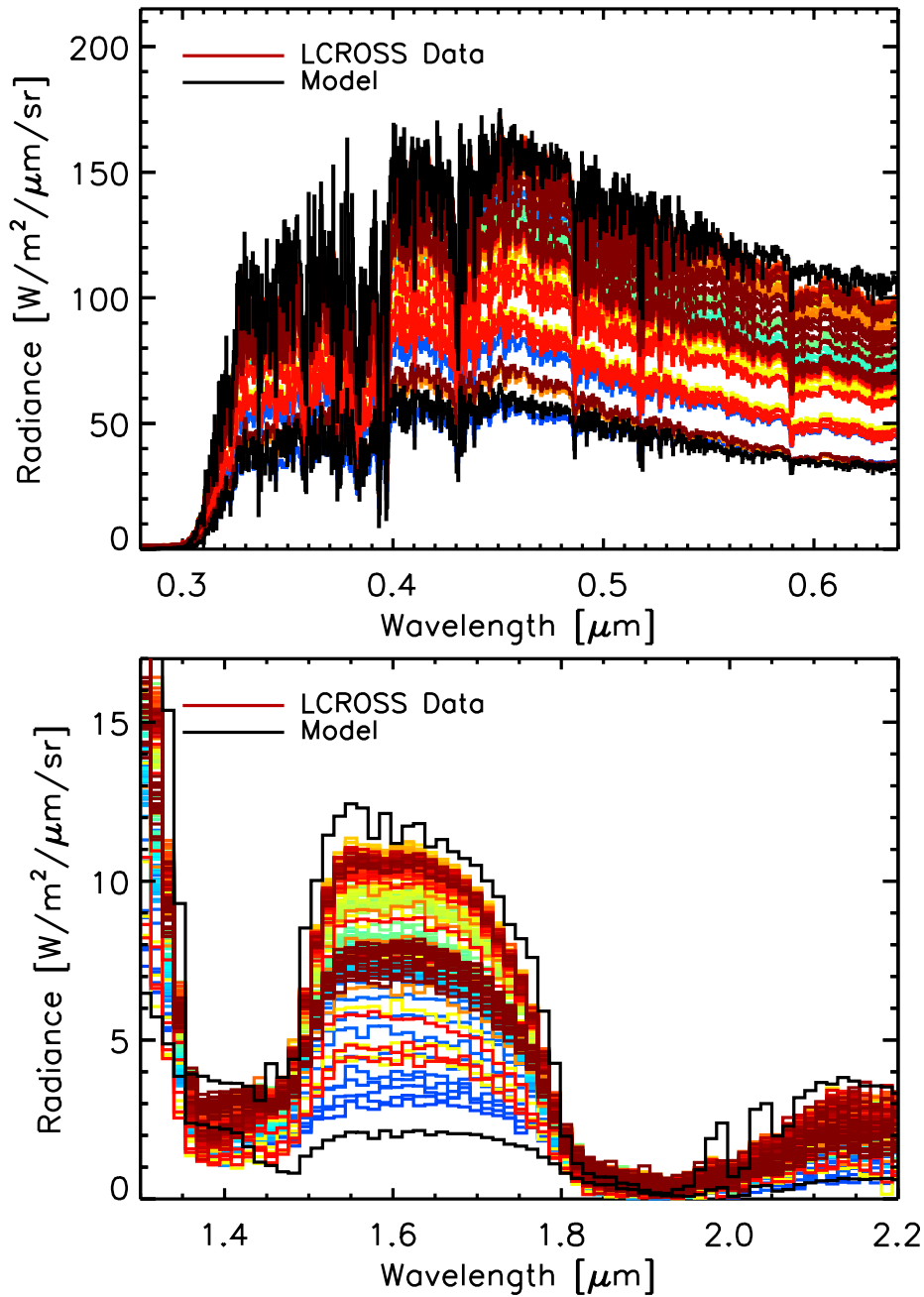

**Figure 7**. Observed and simulated UV-visible (top) and NIR (bottom) spectra of Earth from Earthlook 1. Data are in a variety of colors, which correspond to those in Figure 6. Model spectra are in black and investigate the brightness extremes by using the pointings from the dimmest and brightest observed spectra. The data are for all pointings where the boresight was on Earth's disk, and include 110 UV-visible spectra and 373 NIR spectra.

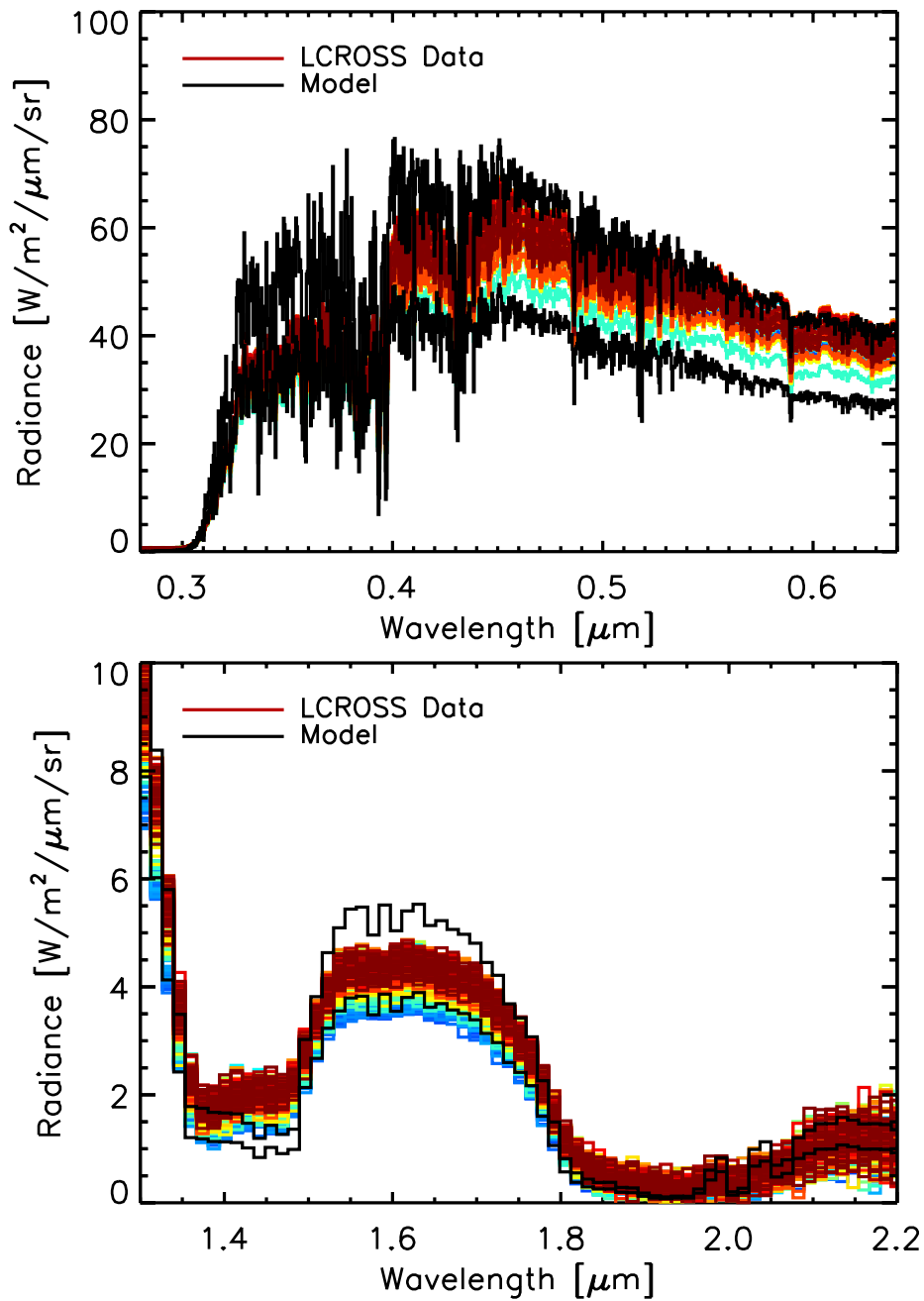

**Figure 8.** Same as Figure 7 except for Earthlook 3. The data are for all pointings where the boresight was on Earth's disk, and include 62 LCROSS UV-visible spectra and 394 NIR spectra.

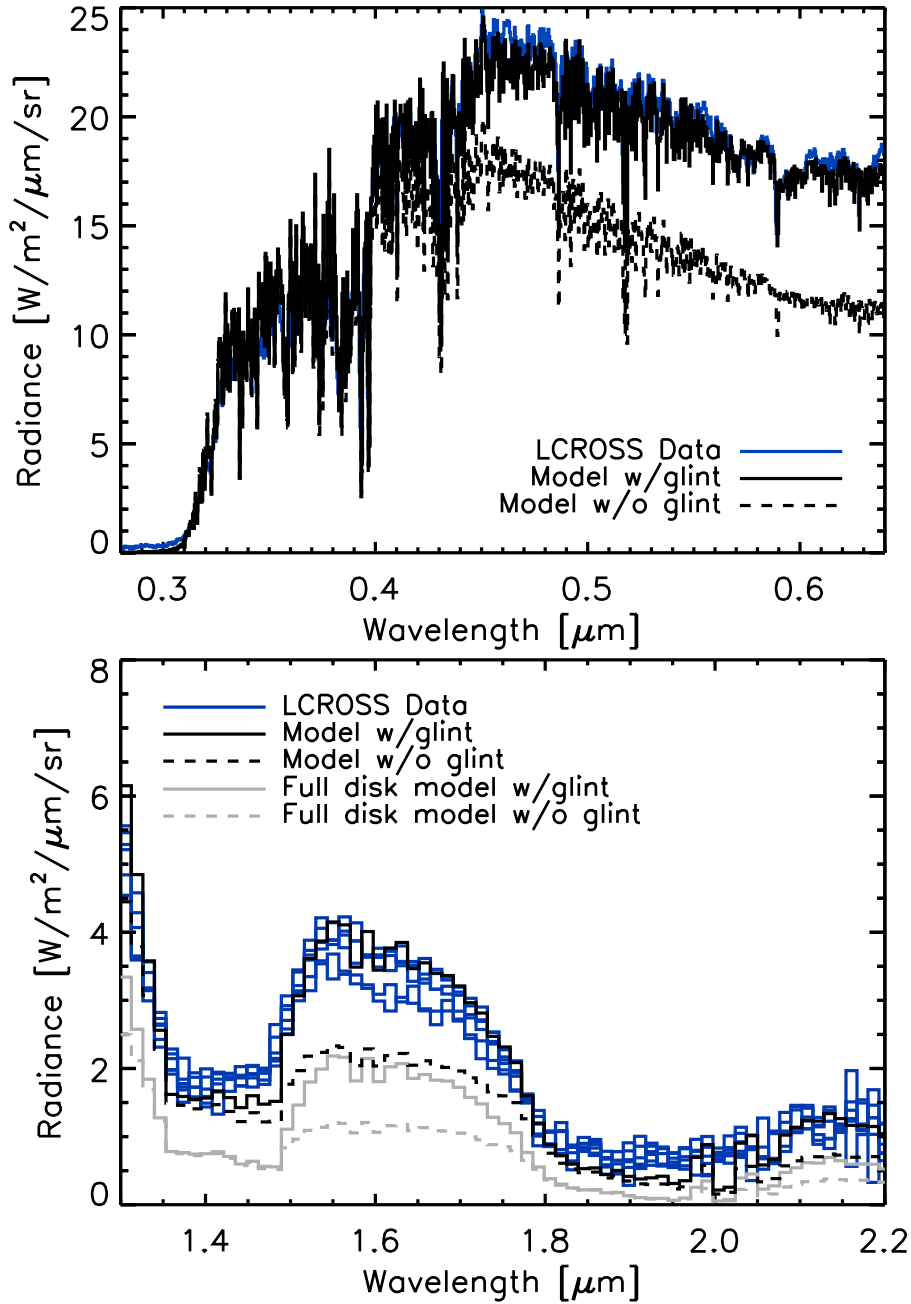

**Figure 9.** Same as Figure 7 except for Earthlook 2. Model brightness extremes are not shown due to the limited number of observations. Instead, models with ocean glint (black, solid) and without glint (black, dashed) are shown. Here we only show the observations that clearly captured the illuminated crescent, which include one UV-visible spectrum and six NIR spectra (see furthest west blue cross in Earthlook 2 image in Figure 6, located in the bright glint spot). Full disk models are shown in the NIR, where the overall dimming is due to integration over the unilluminated portion of the disk.

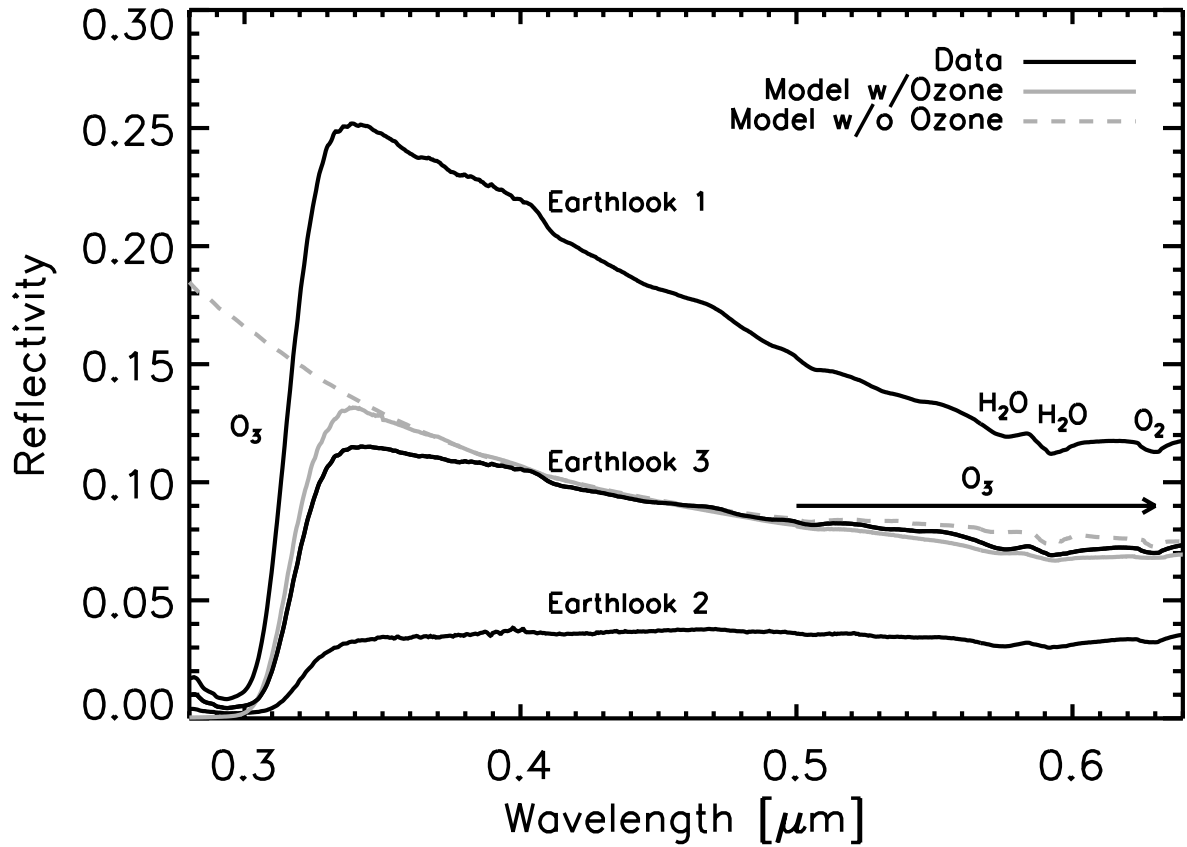

**Figure 10.** Reflectivity of Earth at UV-visible wavelengths. Curves are for intermediate-brightness cases from each Earthlook, and are taken as the intensity measured at the spacecraft divided by the top-of-atmosphere solar intensity, given as $F_\lambda^\odot/\pi$. No corrections for Earth's phase angle have been applied. Notable absorption features are labeled, and a smoothing has been applied to reduce noise. Gray curves are from models for Earthlook 3 both with and without ozone, and show the impact of ozone on the spectra.

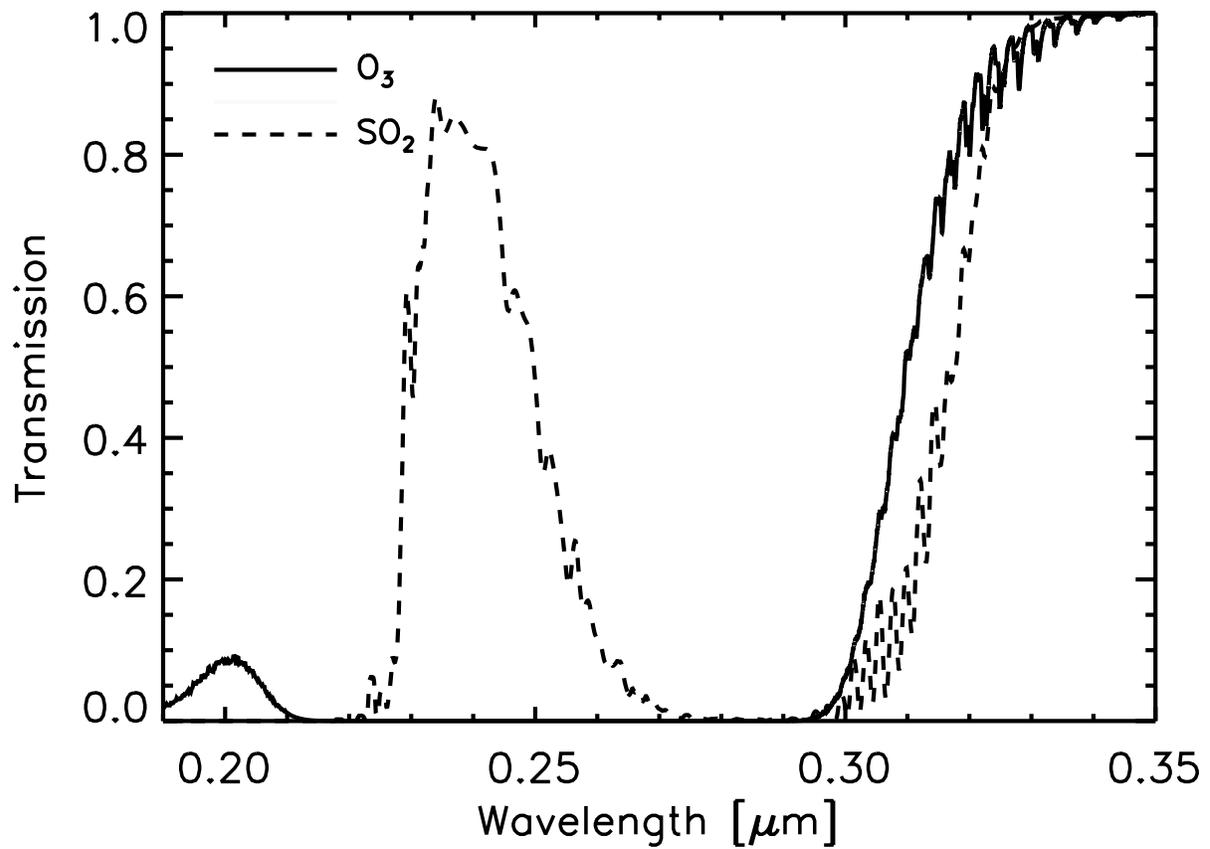

**Figure 11.** Transmission through columns of pure ozone ($O_3$) and pure sulfur dioxide ($SO_2$) at UV wavelengths. Both cases assume a column density of $8 \times 10^{18}$ molecules cm$^{-2}$, which is appropriate for Earth's ozone levels and is about 100 times the column density of $SO_2$ above the Venus cloud tops (Marcq et al. 2012).

TABLES

**Table 1.** Summary of Earth Observations by *LCROSS*

|  | Date | Start Time [UTC] | End Time [UTC] | Earth Phase Angle | Distance from Earth [km] | Earth Angular Diameter | Sub-Spacecraft Lat/Lon | Scene |
|---|---|---|---|---|---|---|---|---|
| Earthlook 1 | 2009-08-01 | 21:51 | 22:48 | 23° | 360,000 | 2.2° | 39°N, 159–164°W | Mid-Pacific, NE Asia, N America |
| Earthlook 2 | 2009-08-17 | 07:52 | 08:27 | 129° | 520,000 | 1.6° | 26°S, 169–179°W | Indian Ocean, South Pacific, Australia |
| Earthlook 3 | 2009-09-18 | 18:08 | 18:46 | 75° | 560,000 | 1.5° | 11°S, 18–28°W | Mid-Atlantic, South America |

**Table 2.** Summary of *LCROSS* Science Instruments used in Earth Observations

| Cameras |  | Pixel FOV [deg] | FOV [deg] | Filter |
|---|---|---|---|---|
| VIS | Visible Camera | 0.043 x 0.047 | 30.5 x 22.8 | CYGM |
| NIR1 | NIR Camera (Filtered) | 0.043 x 0.047 | 28.3 x 21.4 | 1.4–1.7 μm |
| NIR2 | NIR Camera (Un-filtered) | 0.040 x 0.045 | 28.3 x 21.4 | 0.9–1.7 μm |
| MIR1 | MIR Camera (Filtered) | 0.097 x 0.097 | 15.5 x 11.6 | 6.0–10 μm |
| MIR2 | MIR Camera (Un-filtered) | 0.097 x 0.097 | 15.5 x 11.6 | 6.0–13.5 μm |

|  | Spectrometers | FOV (diam.) | Wavelength Range | Resolution (at wavelength) | | |
|---|---|---|---|---|---|---|
| VSP | UV-Visible Spectrometer | 0.98° | 260–650 nm | 1.03 nm (350 nm) | 0.98 nm (450 nm) | 0.66 nm (550 nm) |
| NSP1 | Nadir NIR Spectrometer | 1.01° | 1.17–2.48 μm | 0.0341 μm (1.48 μm) |  | 0.0362 μm (2.06 μm) |

**Table 3.** Number of valid spectra/images taken with Earth in FOV

|  | VSP | NSP1 | NIR1 | NIR2 | MIR1 | MIR2 |
|---|---|---|---|---|---|---|
| Earthlook 1 | 110[a] | 373 | 17 | 10 | 15[d] | 14 |
| Earthlook 2 | 14[a] 14[b] 14[c] | 185 | 18 | 12 | 25 | 25 |
| Earthlook 3 | 62[a] 62[b] | 394 | 30 | 18 | 46 | 46 |

[a]Number of non-saturated 100ms exposure spectra. [b]Number of non-saturated 500ms exposure spectra.
[c]Number of non-saturated 2500ms spectra. [d]Incorrectly taken in low-gain.

**Table 4.** Geometry of *LCROSS* Lunar Swingbys Used for VSP Recalibration

|  | Spacecraft Lat/Lon | Boresight Lat/Lon | Sun Lat/Lon |
|---|---|---|---|
| Swingby 1 (Mendeleev Crater) | 31°N, 271°W | 4.0°N, 216°W | 0.8°N, 170°W |
| Swingby 2 (Goddard Crater) | 34°N, 271°W | 18.4°N, 273°W | 0.8°N, 170°W |
| Swingby 3 (Giordano Bruno Crater) | 36°N, 271°W | 38.4°N, 254°W | 0.8°N, 170°W |